\documentstyle[12pt,cmb,psfig]{article}
\def\beq{\begin{equation}}
\def\eeq{\end{equation}}
\def\bei{\begin{itemize}}
\def\eei{\end{itemize}}

\def\et{{\it et al.\ }}

\def\het{$^{3}$He}

\def\max{MAXIMA}
\def\boom{BOOMERanG }

\def\ghz{GHz}
\def\mkrtsec{$\mu$K$\sqrt{\mbox{sec}}$ }

\def\mathrelfun#1#2{\lower3.6pt\vbox{\baselineskip0pt\lineskip.9pt
  \ialign{$\mathsurround=0pt#1\hfil##\hfil$\crcr#2\crcr\sim\crcr}}}
\def\simlt{\mathrel{\mathpalette\mathrelfun <}}

\begin{document}
\heading{THE MAX AND MAXIMA EXPERIMENTS}

\author{S. Hanany $^{1,2}$, and the MAX$^{1,2,3}$ and MAXIMA$^{1,2,4,5,6,7}$ 
Collaborations} 
   {$^{1}$ The Center for Particle Astrophysics, University of 
California, Berkeley.}  
   {$^{2}$ University of California, Berkeley.}
   {$^{3}$ University of California, Santa Barbara.}
   {$^{4}$ University of Rome.}
   {$^{5}$ IROE-Firenze.}
   {$^{6}$ California Institute of Technology.}  
   {$^{7}$ Queen Marry and Westfield College.}

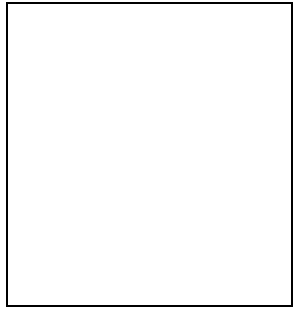
\begin{abstract}{\baselineskip 0.4cm 
We summarize the performance of the MAX experiment during its 5 years
of operation and present a compilation of the results to date.  We
describe MAXIMA, a balloon borne experiment employing an array of
detectors in the focal plane, that will provide sensitive measurements
of the power spectrum between $l \sim 60$ and $l \sim 650$.  }
\end{abstract}

\section{MAX} 


The Millimeter wave Anisotropy eXperiment (MAX) was a balloon borne
experiment that measured the cosmic microwave background anisotropy
(CMBA) on half degree angular scale from 1989 to 1994.  
It was a collaboration between groups in the
University of California, Santa Barbara and Berkeley.  Between 1989
and 1994 the instrument was launched 5 times and scanned 9 regions of
the sky for CMBA.  First detection of CMBA signals was reported by
Alsop \et \cite{alsopetal92}.  
Table 1 summarizes the
flights, the regions scanned, flat band power results, and CMBA
papers published.

\subsection{Overview of the MAX experiment}
\label{The MAX Instrument, Observing Strategy and Window Function}

Several papers describe the MAX instrument 
\cite{alsopetal92,fischeretal92,meinhodletal93b}, here
we only summarize selected aspects of the experiment. Particular
details of the experiment, e.g. exact beam size, frequency bands, and
bolometer temperature, where modified during the duration of the
program. The numbers that will be quoted here refer to the last flight
of the program, MAX-5, unless otherwise noted.


MAX had a single pixel photometer at the focal plane of an off-axis
Gregorian telescope. The telescope provided a beam of 0.5 degree
FWHM. The beam was split inside the photometer to 4 frequency
bands by means of dichroic mesh-filters.

During an observation the beam was modulated on the sky with two
frequencies.  The secondary mirror modulated the beam sinusoidally, in
cross-elevation direction, at a frequency of 5.4 Hz and amplitude of
1.4 deg.  Simultaneously the entire gondola was scanned in azimuth at
constant velocity at an amplitude of 4 degrees and a frequency of
0.0075 Hz.  The center of the gondola scan tracked the location of a
bright star for the duration of the observation.  The fast secondary
mirror chop provided effective discrimination against low frequency
electronic noise which had a $1/f$ knee at $\sim 3$ Hz.  The slow scan
enabled the subtraction of temporal variations in the
bolometer temperature and the atmosphere brightness.

The MAX detectors were composite bolometers operating at 300 mK for the 
first three flights and at 85 mK for the last two. 
The MAX-5 bolometer for the 450 GHz band was background
limited. At the lower frequency bands phonon and Johnson noise was dominant. 
During the duration of the program sensitivity to CMB
temperature differences improved significantly. For example, the 
sensitivity of the 180 GHz channel improved by a factor of 10 from 
$\sim 2$ mK$\sqrt{\mbox{sec}}$ \cite{fischeretal92} 
to 0.24 mK$\sqrt{\mbox{sec}}$ \cite{limetal96}.

The optical chop at $\sim 5$ Hz, which was chosen as an optimum in the 
trade-off between 
the bolometer time constants and the onset of low frequency noise,
determined MAX's $l$ space coverage to a single
window function. The window function peaked at $l = 150$ and had 
half power points at $l= 72$ and $l=248$. 

MAX was calibrated by observing a planet once during the
flight and then using a partially reflecting membrane as a transfer
standard for additional periodic calibrations. The typical absolute
calibration error was 10\% which was dominated by uncertainties in the
brightness temperature of the planets observed.

\subsection{MAX results}

MAX provided seven detections of CMBA signals at an angular scale of
$\sim 0.5$ degrees.  For these seven detections the wide frequency
coverage, up to 4 channels between 90 \ghz\ and 450 \ghz, enabled
unambiguous spectral discrimination against emission from galactic
dust. Extrapolation of the fluctuations observed in the 408 MHz
Haslam map to the MAX frequency bands, 
using the expected spectral dependence of either 
synchrotron or Bremstrahlung radiation, yields a fluctuations'
amplitude of typically less than 10\% of the amplitude observed. Thus
it is unlikely that synchrotron or Bremstrahlung are the dominant source
of the detected fluctuations.    
Searches in available catalogs found no sufficiently intense radio
sources in the regions observed.  The treatment of potential temporal
variations in the signal due to atmosphere variability, beam motion
relative to the balloon or earth, moon location, etc. are discussed in
the references mentioned in Table 1.  Two measurements near the star
$\mu$-Pegasi, where 100$\mu$ IRAS maps indicate significant dust
contrast, were expected to reveal dust signals. Indeed, the dust
signature detected was morphologically consistent with IRAS. 
Only upper limits on the CMB fluctuation power were derived in these
regions.

Most cosmological models predict an increase in the power spectrum of
the CMB fluctuations near the peak of MAX's window function. The MAX
results are suggestive of a combined flat band power larger than that
detected by the COBE/DMR experiment. Statistical analysis to combine
the seven detections to a single estimate of the CMB fluctuation power
within MAX's window function is in progress.

\begin{table}[htb]
\label{table max summary}
\centerline{ \begin{tabular}{|| l | c | l ||} \hline
Flight/Year/ Region Observed & $(\Delta T/T)_{flat}$ & Publication 
 \\ \hline \hline
MAX--1 / 1989  &              &  Fischer \et 1992 \cite{fischeretal92}  
 \\ \hline 
MAX--2 / 1990 / GUM$^{*}$ & $2.9^{+4.3}_{-1.8}$ & Alsop \et 1992 \cite{alsopetal92}  
\\ \hline 
MAX--3 / 1991 / GUM$^{*}$ & $2.7^{+1.1}_{-0.7}$ & Gundersen \et 1993 
\cite{gundersenetal93}  \\  \hline
MAX--3 / 1991 /$\mu$-Pegasi & $<$ 1.6 & Meinhold \et 1993 
\cite{meinholdetal93} \\  \hline 
MAX--4 / 1993 / GUM$^{*}$ & $2.0^{+0.6}_{-0.4}$$^{**}$ & Devlin \et 1994 
\cite{devlinetal94}  \\  \hline 
MAX--4 / 1993 / $\sigma$-Herculis & $1.8^{+0.8}_{-0.6}$$^{**}$ & 
Clapp \et 1994 \cite{clappetal94} \\  \hline 
MAX--4 / 1993 / $\iota$-Draconis & $1.9^{+0.7}_{-0.4}$$^{**}$ & 
Clapp \et 1994 \cite{clappetal94} \\   \hline
MAX--5 / 1994 / HR5127 & $1.2^{+0.4}_{-0.3}$  & Tanaka \et 1996
\cite{tanakaetal96}  \\  \hline 
MAX--5 / 1994 / $\phi$-Herculis & $1.9^{+0.7}_{-0.4}$  & Tanaka \et 1996 
\cite{tanakaetal96} \\  \hline
MAX--5 / 1994 / $\mu$-Pegasi & $ < 1.3 $ & Lim \et 1996 \cite{limetal96} 
 \\  \hline
\end{tabular} }
\caption{Summary of MAX results. Values of $\Delta T /T$ are for 
flat band $\langle {l(l+1) C_{l} \over 2\pi} \rangle^{1/2} $, 
95\% confidence interval. ($^{*}$) GUM stands for the region near the 
star Gamma Ursa Minoris. ($^{**}$) Original results were revised 
as described by Tanaka \et (1996). }
\end{table}

\section{MAXIMA}

The goal of next generation experiments is to make precise
measurements of the CMBA power spectrum.  Theoretical work
within the last several years has demonstrated that the optimal 
observing strategy to constrain
the power spectrum, in the absence of systematic errors or
foregrounds, is to observe as many sky
pixels as possible with modest ($\sim 1$) signal to noise per pixel
\cite{knox95}. It has also been argued that small to intermediate
scale measurements, at $100 \simlt l \simlt 1500$ covering the region
where CDM models predict adiabatic peaks, could discriminate between
various cosmological models \cite{whitehumoriond,joaoalbrecht96} and
provide information about the cosmological parameters independent of
the underlying cosmological model \cite{huwhitemoriond}.  

The Millimeter wave Anisotropy eXperiment Imaging Array (MAXIMA) was
designed to address these scientific requirements by scanning many
pixels on the sky within a single flight, providing large $l$ space
coverage and high $l$ resolution, while improving on the systematic-error
rejection achieved for MAX.
\max\ is a balloon borne program designed to constrain the 
CMBA power spectrum on a range of angular scales between $l \sim 60$
and  $l \sim 650$.  It is a collaboration between groups at the University of
California, Berkeley, University of Rome, IROE -- Florence, Queen
Mary and Westfield College -- London, and the California Institute of
Technology.

\subsection{Experimental Configuration}

MAXIMA will observe 14 sky pixels simultaneously with 0.18 degree
FWHM beams.  The attached Figure shows the experiment, the focal plane
and its orientation on the sky. The 14 single
frequency photometers detect radiation in three frequency
bands centered around 150 GHz, 240 GHz, and 420 GHz. The bolometers
will be maintained at $100$ mK to provide high sensitivity and short
time constants. The experiment is designed for up to 24 hour of
observations and it will fly in north America.

\begin{figure}
\centerline{\psfig{file=cryosta4.plt,height=7.5in}}
\label{figure maxima experiment}
\end{figure}

\subsubsection{optics}

The optical system is a three mirror off-axis f/1.8 Gregorian
telescope.  The primary mirror is a $1.3$ meter diameter off axis section of
a parabola.  The secondary and tertiary mirrors (21 cm and 18 cm in
diameter respectively) are conic sections with aspheric components
which compensate the aberrations introduced by the primary mirror.
The secondary and tertiary mirrors and a Lyot stop are mounted in a
well baffled box inside the cryostat and are cooled to liquid helium
temperature. The cold Lyot stop provides excellent sidelobe rejection
and cooling the secondary optics
reduces the optical loading on the bolometers. A three
mirror system was designed to provide for a diffraction limited $\sim
1 \times 1$ deg$^2$ field of view at 150 GHz.  The
secondary and tertiary mirrors are fixed and the light, 11 kg, primary
mirror can be modulated around the optical axis of the telescope (the
line connecting the center of the primary and the center of the
secondary).

\subsubsection{Cryogenics, Detectors and Electronics}

The cryostat was designed for a north-American flight of up to 24
hours.  
The bolometers will be cooled to 100 mK by means of an
adiabatic demagnetization refrigerator (ADR).  
The heat of magnetization generated during the ADR cycle
will be sunk into a \het\ refrigerator operating at 300 mK with 
a cooling capacity of 25 Joules. The resulting 
cooling capacity of the ADR is 93 mJoules. With expected heat loads 
both the ADR and the \het\ refrigerator will maintain cooling capacity
much longer than the cryostat. 
 
Spider-web bolometers \cite{bock94} operating at 100 mK will be used to
detect the incoming radiation\footnote{See also a paper by Debernardis in 
these proceedings. 
The bolometers, readout electronics, and attitude control system are shared
technology between \boom\ and \max.}. 
Extrapolation from measurements at 300 mK, and preliminary 
measurements at 100 mK, indicate that the detectors
will be background limited and will have time constants $\simlt 10$ msec. 
We expect a detector NET of 60 \mkrtsec (90 \mkrtsec) at the 
150 GHz (240 GHz) frequency band. 

The detectors will be AC-biased at a frequency of several hundred
Hz$^{1}$. The post lock-in noise of the readout electronics 
was measured to be less than 10 $n$V$\sqrt{\mbox{Hz}}$, 
down to frequencies smaller than 100 mHz.  

\subsubsection{Gondola and Attitude Control}

The gondola provides for pointing in azimuth and elevation.  Pointing
control is achieved with a 5 Hz feedback loop control relying on a two
axis magnetometer for coarse pointing ($\pm 2$ degrees) and on a CCD
camera as a fine sensor. The CCD camera and its associated f/0.7 lens
provide a field of view of 7.4 degrees in azimuth and 5.5
degrees in elevation, and pixel resolution of 0.8 arcmin/pixel and
0.9 arcmin/pixel, respectively.  The on board image processing is
expected to provide sub-pixel resolution.
Overall pointing stability is expected to be 1 arcminute RMS or better.

\subsection{Observing Strategy and $l$ Space Coverage}
\label{Observing Strategy and $l$ Space Coverage}

\max's beam will be scanned in azimuth with two frequencies. The primary 
mirror will modulate the beam in a triangular wave with frequency $f_{1}$
and amplitude $A_{1}$, while the gondola will be simultaneously scanned 
in azimuth at a slower rate. Here we discuss the choice of $f_{1}$ and 
$A_{1}$. 

A bolometer with time constant $\tau$ acts as a single pole low pass
filter on the detected optical signals. The -3 dB point of this filter
is used to set a criterion on the maximum speed that the beam can be
scanned across the sky. The signal detected by a fast bolometer ($\tau
\simeq 0$) when a Gaussian beam with width $\sigma = 0.425 \times
\mbox{FWHM}$ crosses a point source at constant speed $\dot{\theta}$
has a Gaussian frequency distribution with width
$\tilde{\sigma}=\dot{\theta}/(2 \pi \sigma)$. If we require that the
-3 dB roll-off of a real bolometer will be larger than
$3\tilde{\sigma}$ we obtain a relation between the maximum scan speed
and the bolometer time constant
\beq
{1 \over 2 \pi \tau} \geq { 3 \dot{\theta} \over 2 \pi \sigma} 
\; \; \Rightarrow \; \;  
\dot{\theta} \leq {\sigma \over 3 \tau} = {\mbox{beam FWHM} \over 7 \tau}.
\eeq 
For a 0.18 degrees FWHM beam width and $\tau=10$ msec, 
$\dot{\theta} \leq 2.6$ degrees/sec. By moving the beams across
the sky at this (or somewhat lower) speed the bolometers remain
sufficiently sensitive to all spatial frequencies up to $\sim$1/beam size. 

The amplitude $A_{1}$ is determined by requiring that the scan 
frequency $f_{1}$ be higher than the knee of the $1/f$ noise. 
preliminary measurements during the first flight of \max\ suggest that 
$f_{1} \simeq 0.5$ Hz is appropriate. For a 
triangular wave 
\beq 
4 A_{1} f_{1} = \dot{\theta} \leq 2.6 \; \mbox{deg/sec},
\eeq
so that $A_{1} = 1.3$ degrees. Larger amplitudes are possible with shorter
bolometer time constants. 

This scan strategy is efficient and enables the synthesis
of multiple window functions in a single scan. 
In combination with the 11 arcminute beams we expect 
an $l$ space coverage between $l=60$ and $l=650$. (see also a 
companion paper in this proceedings \cite{adrianmoriond}.)  



\subsection{Status and Flight Program}

The \max\ set of measurements is being comissioned in stages.  In the
first flight, which was launched from Palestine, Texas, on Sept. 2,
1995, we flew the single beam receiver used on MAX-4, and MAX-5. Most
other flight systems, including the gondola, the pointing system,
AC-bias electronics, and chopping primary mirror, were new. The flight
goals were to test all new flight systems, scan regions of the
sky for CMBA signals, and test new scan strategies.  All of these
have been successfully accomplished. Data analysis is in progress.

The 14-beam array is presently under construction and 
is scheduled to be launched as \max-2 in the spring of 1997. 


%
\vfill

\begin{thebibliography}{99}{\baselineskip 0.4cm

\bibitem{alsopetal92}
Alsop, D. \et 1992, ApJLett, 395, 317

\vspace{-0.1in}
\bibitem{fischeretal92}
Fischer, M. L. \et 1992, ApJ, 388, 242

\vspace{-0.1in}
\bibitem{meinhodletal93b}
Meinhold, P., \et 1993, ApJ, 406, 12

\vspace{-0.1in}
\bibitem{gundersenetal93}
Gundersen, J. O. \et 1993, ApJ, 413, L1 

\vspace{-0.1in}
\bibitem{meinholdetal93}
Meinhold, P., \et 1993, ApJ, 409, L1

\vspace{-0.1in}
\bibitem{devlinetal94}
Devlin, M., et al. 1994, ApJ, 430, L1

\vspace{-0.1in}
\bibitem{clappetal94}
Clapp, A. C., \et 1994, ApJ, 433, L57

\vspace{-0.1in}
\bibitem{tanakaetal96}
Tanaka, S. T., \et 1996, ApJ, in publication

\vspace{-0.1in}
\bibitem{limetal96}
Lim, M. A., \et 1996, ApJ, in publication

\vspace{-0.1in}
\bibitem{knox95}
Knox, L. 1995, Phys. Rev. D, 52, 4307 

\vspace{-0.1in}
\bibitem{whitehumoriond}
M. White, \& W. Hu 1996, these proceedings

\vspace{-0.1in}
\bibitem{joaoalbrecht96}
M. Magueijo, A. Albrecht, D. Coulson, \& P. Ferreira 1996, Phys. Rev. Lett, 76, 
2617

\vspace{-0.1in}
\bibitem{huwhitemoriond}
W. Hu, \& M. White 1996, these proceedings

\vspace{-0.1in}
\bibitem{jungmankam95}
Jungman, G., Kamionkowski, M., Kosowsky, A., \& Spergel, D. N. 1996,
Phys. Rev. D, in press



\vspace{-0.1in}
\bibitem{bock94}
Bock, J. J., Chen, D., Mauskopf, P. D., \& Lange, A. E. 1994, Proceedings
of The Future of IR and MM-Wave Astronomy, Saclay, France.  

\vspace{-0.1in}
\bibitem{adrianmoriond}
Lee, A. T., \et 1996, these proceedings. }

\end{thebibliography}
\end{document}